\documentclass[amsmath,aps,prb,twocolumn]{revtex4}
\usepackage{amsthm,amsfonts,graphicx,verbatim,color,cancel,amsmath,caption,subcaption}
\newcommand{\mathsym}[1]{{}}
\newcommand{\unicode}[1]{{}}

\newcommand{\bea}{\begin{eqnarray}}
\newcommand{\eea}{\end{eqnarray}}

\newcommand{\ket}[1]{\ensuremath{\left\vert #1 \right\rangle}}
\newcommand{\bra}[1]{\ensuremath{\left\langle #1 \right\vert}}

\newcommand{\absval}[1]{\ensuremath{\left \vert #1 \right\vert}}

\newcommand{\bk}{\ensuremath{\mathbf{k}}}
\newcommand{\bQ}{\ensuremath{\mathbf{Q}}}

\begin{document}

\title{Auxiliary fermion approach to the RIXS spectrum in a doped cuprate}

\author{Yifei Shi$^{1,2}$, Andrew J. A. James$^{3}$, Eugene Demler$^{4}$ and Israel Klich$^{1}$}

\affiliation{$^{1}$ Department of Physics, University of Virginia, Charlottesville, VA 22904, USA\\ $^{2}$ {Department of Chemistry, McGill University, Montr{\'e}al, Qu{\'e}bec H3A 0B8, Canada\\}$^{3}$ London Centre for Nanotechnology, University College London, Gordon Street, London WC1H 0AH, United Kingdom\\ $^{4}$ Physics Department, Harvard University, Cambridge, Massachusetts 02138, USA}

\begin{abstract}
%We describe a method for calculating the resonant inelastic X-ray scattering (RIXS) response---including transient core hole---of many-body systems with certain non-trivial effects encoded in their single particle propagator.
We describe a method for calculating the resonant inelastic X-ray scattering (RIXS) response---including the dynamics of the transient core hole---of many-body systems with non-trivial gap structure encoded in their single particle Green's function.
Our approach introduces auxiliary fermions in order to obtain a form amenable to the determinant method of Benjamin {\it et al.}\cite{Benjamin2014}, and is applicable to systems where interactions are most strongly felt through a renormalization of the single particle propagator.
As a test case we consider the Yang Rice Zhang ansatz describing cuprate phenomena in the underdoped `pseudogap' regime, and show that including the core hole dynamics pushes the RIXS peaks towards higher energy transfer, improving agreement with experiments.

%We show how to adapt an exact treatment of the core hole in resonant inelastic X-ray scattering (RIXS) to also include many-body interaction effects, by introducing an action with auxiliary fermions.
%A phenomenological ansatz for the Green's function of doped cuprates, the Yang-Rice-Zhang (YRZ) Green's function, has been successful in describing the Fermi surface structure and magnetic properties of cuprates. We propose a method for using this Green's function as the underlying action in dynamical processes such as resonant inelastic X-ray scattering (RIXS), by introducing auxiliary fermions. This approach allows an exact calculation of the RIXS spectrum using a recent method that takes into account the effect of the core hole, which is otherwise difficult. The core hole pushes the RIXS peaks towards higher energy transfer, improving agreement with experiments. 
\end{abstract}

\maketitle
\section{Introduction}
In order to assess the validity of theories of strongly correlated phenomena in materials, it is necessary to compare data from experimental probes to theoretical predictions.
If these comparisons are to be meaningful, it is vital to accurately account for the physics of the measurement techniques.
Doing so for inelastic neutron scattering studies of quantum spin chains has enabled highly sensitive experimental tests, in some cases providing excellent support for theory\cite{lake2013multispinon}.
Conversely, for the cuprate high temperature superconductors the origins and roles of many observed features remain unclear, despite a wealth of experimental studies.
It is therefore of paramount importance to be able to discriminate between the experimentally observable aspects of different theories of cuprate phenomena.

Recently, resonant inelastic X-ray scattering (RIXS) has emerged as a very useful probe of strongly correlated condensed matter \cite{dean2015insights,vernay2008cu,chen2010unraveling,Guarise2014cuprate}.
One of the exciting features of RIXS compared to other probes, such as angle-resolved photoemission spectroscopy (ARPES) or neutron scattering, is the ability to reach high energy and momentum transfer, which makes it possible to probe the full dispersion of excitations \cite{ament2011resonant}.
RIXS may also serve as a sensitive measure of band structure both below and above the Fermi level in itinerant electron systems \cite{kanasz2016resonant}.

A subtle point in understanding RIXS is the precise role of the core hole potential during the intermediate states of the scattering process.
Often it is an excellent approximation to assume a very short life time for the hole, allowing one to incorporate the hole in a relatively straight forward manner and relate the signal to a dynamical susceptibility function.
However, in some systems a more sophisticated treatment is necessary.
For example, a recent comparative study employing exact diagonalization shows that the ultra short life time core approximation is not always enough to describe the RIXS signal in a qualitatively accurate way, because of dynamics associated with the presence of the core hole \cite{jia2016using}.
%shift the shape of dispersion curves, and lead to significant effects when one compares the so called spin flip and non-spin flip scattering channels \cite{Benjamin2014}.
Disentangling these `experimental' effects from those originating in different theoretical descriptions of the system under study will only become more important in future as RIXS resolution improves.

An analytical treatment of the RIXS core hole---applicable to simple hopping hamiltonians, including possible pairing terms---suggests that it can have significant effects for band structures with parameters appropriate to the cuprates \cite{Benjamin2014,kanasz2016resonant,shi2016superconducting}.
Nevertheless, the cuprates are strongly correlated materials, and one must go beyond simple band structure models to capture their most interesting properties.

Here we propose a method that extends the treatment of the RIXS core hole due to Benjamin {\it et al.}\cite{Benjamin2014}, to non-trivial systems with interaction effects encoded in their single particle Green's function.
To demonstrate our approach, we consider the Yang-Rice-Zhang (YRZ) ansatz \cite{YRZ} for cuprates in the `pseudogap' region, which takes the form of a phenomenological, interacting Green's function.

The pseudogap region epitomises the strongly correlated behaviour of the cuprates and is characterised by an anomalous Fermi surface---between the undoped insulating and heavily overdoped metallic phases---which consists of four disconnected arcs \cite{norman1998destruction}.
Motivated by studies of weakly coupled Hubbard ladders \cite{KRT2006}, Yang, Rice and Zhang\cite{YRZ} (YRZ) proposed a phenomenological ansatz Green's function to describe this peculiar structure.
The YRZ propagator yields a Fermi surface of four hole pockets, with area proportional to doping $x$, and vanishing spectral weight at the backs of the pockets due to lines of `Luttinger zeros'.
This ansatz has proven effective at reproducing and parametrizing the results ARPES \cite{Yang2008} and a variety of other experimental probes \cite{rice2011phenomenological}.
Recently it has been found to be consistent with the particle-hole asymmetric gap detected by time resolved ARPES\cite{miller2017particle}.
We also note that approaches conceptually similar to YRZ, that introduce phenomenologically finite quasiparticle lifetime, have recently been used to describe both ARPES and STM experiments in high-$T_c$ cuprates\cite{vishik2012phase,reber2012origin,dalla2016friedel}.

Initially formulated as a two point function, the YRZ ansatz was extended to describe higher order correlation functions \cite{James2012} by connecting it to a slave boson treatment of the $t-J$ model \cite{brinckmann2001}.
In this form, combined with YRZ band parameters provided by ARPES, it has been useful for interpreting RIXS results \cite{DeanBi2212,Dean2014layer}.
However to date there has been no attempt to systematically incorporate the important physics of the transient core hole potential in the RIXS response predicted by YRZ.
Indeed, since the theory is not free, it does not allow the calculation of some other quantities such as density-density correlations, without making further assumptions \cite{James2012}. 

In the next section we show how to calculate the RIXS response of a system with a YRZ-like Green's function, while still treating the core hole rigorously using the methods Ref.~\onlinecite{Benjamin2014}.
We then compare our results to experiments on the high-$T_{c}$ cuprate Bi-2201 in the hole underdoped regime\cite{Dean2014layer}, and demonstrate that inclusion of the core hole improves on existing calculations by shifting peak positions at higher momenta.

\section{Method}
We will frame our discussion in terms of the YRZ propagator, but note that the method is generally applicable to single particle propagators of the form
\begin{align}
G^{-1}(\omega,\bk)=\omega -E_\bk -\frac{\absval{f_\bk}^2}{\omega+\ell_\bk},
\label{eq:propagator}
\end{align}
where $E_\bk, f_\bk$ and $\ell_\bk$ are all functions of the single particle momentum $\bk$.
Note that this differs from the familiar pairing gap propagator in BCS theory because $E_\bk \ne \ell_\bk$, and instead allows for asymmetry between the bands above and below the gap.
Such propagators admit poles (corresponding to coherent quasiparticle excitations) when $\omega -E_\bk -\absval{f_\bk}^2/(\omega+\ell_\bk)=0$ and zeros when $\omega+\ell_\bk=0$.

\subsection{The YRZ propagator}
The YRZ ansatz takes the coherent part of the electron Green's function in a two dimensional copper oxide plane, independent of spin, as\cite{YRZ}:
\begin{align}
G_{e}(\omega,\bk)=\frac{{g_t(x)}}{\omega-\xi_{0}(\bk)-\xi'(\bk)-\frac{{|\Delta_{\text{RVB}}(\bk)|^{2}}}{\omega+\xi_{0}(\bk)}}.
\label{eq:YRZ}
\end{align}
Here, $x$ is the doping, $\xi_0,\xi'$ are bands with `renormalized' parameters, and $\Delta_{\text{RVB}}(\bk)=-\Delta_0(\cos k_{x}-\cos k_{y})$ is a `Resonating Valence Bond' (RVB) gap function.
Throughout the paper we neglect the superconducting gap, which is much smaller than all other parameters we consider, and so for notional convenience we drop the RVB subscripts in the expressions below.
For the bands we take $\xi_{0}(\bk)=-2t(x)(\cos k_{x}+\cos k_{y})$, $\xi^{'}(\bk)=-4t^{'}\cos k_{x}\cos k_{y}-2t^{''}(\cos 2k_{x}+\cos 2k_{y})-\mu_{p}$.
The hopping parameters depend on doping as: $t(x)=g_{t}(x)t_{0}+\frac{3}{8}g_{s}(x)J_H\chi$, $t^{'}(x)=g_{t}(x)t_{0}^{'}$, $t^{''}(x)=g_{t}(x)t_{0}^{''}$, $\chi \sim \langle a^{\dagger}_{i\sigma}a_{i+\hat{x},\sigma}\rangle$.
$g_t,g_s$ are referred to as the `Gutzwiller functions' \cite{YRZ}, and are given here by  $g_t  = \frac{2x}{1+x}$, $g_s = \frac{4}{(1+x)^2}$.
$\mu_p$ is a chemical potential term that is determined by the doping \cite{Tsvelik2006}.
The inclusion of $g_t(x)$ in the numerator reflects the overall weight of the coherent part of the single particle propagator, relative to the incoherent part.

The Green's function, Eq. \ref{eq:YRZ}, can also be viewed as the result of a slave boson, renormalized mean field theory, treatment of the $t\text{-}J$ model\cite{brinckmann2001}, with a particular resummation of the hopping terms\cite{James2012}.
In this picture the electron is factored into fermionic spinon and bosonic holon degrees of freedom, and the assumed condensation of the latter yields the $g_t(x)$ factor.
The spinon Green's function is then
\begin{align}
G_{s}(\omega,\bk)=\frac{1}{\omega-\xi_{0}(\bk)-\xi'(\bk)-\frac{{|\Delta(\bk)|^{2}}}{\omega+\xi_{0}(\bk)}}.
\label{eq:spinon}
\end{align}

Because the YRZ model has many parameters, a free fit to the RIXS data would have little edifying value.
Instead we follow Refs.~\onlinecite{DeanBi2212,Dean2014layer} and use parameters independently determined by ARPES \cite{Yang2008}.
This will also allow us to make a direct comparison with preexisting calculations of the RIXS response.
For convenience the parameter values we use in the calculations below, with $x=0.12$ (corresponding to an underdoped sample) are shown in Table \ref{param}. 
%\begin{center}
\begin{table}
\caption{YRZ parameters \label{param}}
\begin{tabular}{ c|c|c|c|c|c|c } 
$t_0$  & $t_0^{'}$ & $t_0^{''}$ & $J_H$ & $\chi$ & $\Delta_0$ & $\mu_p$ \\
\hline
0.144eV & -0.3$t_0$ & 0.2$t_0$ & 0.12eV & 0.338 & 0.3$t_0$  & -0.0571eV
\end{tabular}
\end{table}
%\end{center}

\subsection{Action and auxiliary fermions}
We begin by formulating an action for a fermionic field with a Green's function of the YRZ type, Eq. \ref{eq:spinon} (equivalently one could work with the definition in Eq. \ref{eq:propagator}):
%In order to describe the dynamical process involved in a RIXS experiment, it is possible to employ the YRZ Green's function as an effective action for spinons, as:
\begin{align} \label{Somega}
S[{\nu}]_0=\int \!\!\text{d}\omega \sum_\bk {\nu}_\bk (\omega-\xi_{0}(\bk)-\xi'(\bk)-\frac{{|\Delta(\bk)|^{2}}}{\omega+\xi_{0}(\bk)})\bar{{\nu}}_\bk,
\end{align}
We concentrate below on the low temperature limit, $T\rightarrow 0$. Written explicitly in a temporal representation,
\bea\label{spinonAct}
S[{\nu}]_0&=&\frac{1}{2\pi}\int\! \text{d}t \sum_\bk{\nu}_\bk(t)\big(i \partial_t-\xi_{0}(\bk)-\xi'(\bk)\big)\bar{{\nu}}_\bk(t) \nonumber \\
&-& \frac{1}{4\pi^2}\int \!\text{d}t_1 dt_2 \sum_\bk{\nu}_\bk(t_1)\bar{{\nu}}_\bk(t_2)h(t_2-t_1).
\eea
The action \eqref{spinonAct} is non-local in time, with a response kernel: 
\bea 
h(t) = \int_{-\infty}^{\infty}\! \text{d}\omega \frac{\absval{\Delta(\bk)}^2}{\omega+\xi_0(\bk)}e^{i\omega t}.
\eea

In the RIXS procedure, when the X-ray knocks a core electron out and creates a core hole, it generates a temporary local potential (whose duration is decided by the core hole lifetime), this quench-like process is often modeled\cite{van2005theory} as turning on a point interaction potential from time $t=0$ to time $t=\tau_0$.
We note that the while the core hole potential only acts directly on charged particles, in the slave boson version of YRZ the implicit hard core constraint suggests that there will be an effective attractive potential for spinons (since the charged holons are repulsed by the core hole).
%We note that the while the core hole potential acts directly on the charged holon, the hard core constraint implicit in the slave boson formulation suggests that the effective potential will also be present for spinons. {\color{cyan}This effective potential is expected to be attractive for spinons since holons are repelled by the core hole.}

The action including a core hole is $S_{\text{corehole}}=S[{\nu}]+\int_0^{\tau_0} \!\text{d}t U_c{\nu}_r\bar{{\nu}}_r$. 
At this stage, the non-local nature of the action in Eq. \eqref{Somega} makes it awkward to analyze. To deal with this problem, we add an auxiliary fermion $\psi_k$, that reproduces the spinon action \eqref{Somega} for ${\nu}_k$ while retaining a quadratic and time-local form:
\bea\label{Snupsi}
S[{\nu},\psi]_0 &=&  {\frac{1}{2\pi}}\int \!\text{d}t \sum_{\bk} \Big[{\nu}_\bk(t)\big(i \partial_t-\xi_{0}(\bk)-\xi'(\bk)\big)\bar{{\nu}}_\bk(t)\nonumber \\
&+&\psi_\bk(t)\big(i \partial_t+\xi_{0}(\bk)\big)\bar{\psi_\bk}(t)\nonumber \\
&+&\Delta(\bk){\nu}_\bk(t)\bar{\psi}_{-\bk}(t)+\bar{\Delta}(\bk)\psi_{-\bk}(t)\bar{{\nu}}_\bk(t)\Big].
\eea
Integrating out the $\psi$ field would yield the action in Eq. \eqref{Somega}. Notice that $\xi_0(\bold{-k})=\xi_0(\bold{k})$, and that the hopping parameters of the auxiliary fermion $\psi_\bk$ are shifted by $\xi'(\bk)$ compared to $\nu_\bk$.
In Ref.~\onlinecite{sakai2016hidden}  a related hidden fermion representation was recently used for dynamical mean field calculations.

The advantage of this formulation is that we are now in position to easily use the methods of Ref. \onlinecite{Benjamin2014}, since the new action is well described by a tight binding hamiltonian. Including a spin index, our hamiltonian is:
\bea \label{Ham} 
H_{cd}&=&-\sum_{ij,\sigma=\uparrow,\downarrow}t_{ij}^{c}c_{i\sigma}^{\dagger}c_{j\sigma}-\sum_{ij,\sigma=\uparrow,\downarrow}t_{ij}^{d}d_{i\sigma}^{\dagger}d_{j\sigma} \nonumber \\
&+&\sum_{ij,\sigma=\uparrow,\downarrow}\Delta_{ij}c_{i\sigma}^{\dagger}d_{j\sigma}+h.c.
\eea
where the $c_{i\sigma}$ and $d_{i\sigma}$ annihilation operators correspond to the original fermions (spinons in the YRZ ansatz) and the auxiliary fermions respectively. The hopping and pairing parameters are: $t^c_{i,i\pm\hat{x}}=t^c_{i,i\pm\hat{y}}=t$, $t^c_{i,i\pm\hat{x}\pm\hat{y}}=t^{'}$, $t^c_{i,i\pm2\hat{x}}=t^c_{i,i\pm2\hat{y}}=t^{''}$, $t^c_{i,i} = -\mu_p$, $t^d_{i,i\pm\hat{x}}=t^d_{i,i\pm\hat{y}}=-t$, $\Delta_{i,i+\hat{x}}=-\Delta_{i,i+\hat{y}}=\Delta$. $t^c_{ij}$ contains a nearest neighbor hopping, next nearest neighbor hopping and a chemical potential term, and $t^d_{ij}$ only contains a nearest neighbor hopping term, which differs from that in $t^c_{ij}$ by a sign.

\subsection{Dynamical core hole}
We now consider the RIXS response for the system described by the hamiltonian Eq. \eqref{Ham}.
The Kramers-Heisenberg formula (see, e.g. Ref. \onlinecite{ament2011resonant}) for the intensity with photon energy and momentum transfer $\omega \to \omega - \Delta\omega$ and $\bk \to \bk + \bQ$ respectively, is given by:
\begin{align}
\label{K-H}
I(\bold{Q},\Delta\omega)\propto&\sum_{f}| A_{f}|^{2}\delta(E_{f}-E_{i}-\Delta\omega), \nonumber \\
A_{f}=&\sum_{m}e^{i\bQ\cdot\mathbf{R}_m}\chi_{\rho\sigma} \sum_{n}\frac{\bra{f}c_{m\rho}\ket{n}\bra{n}c_{m\sigma}^{\dagger}\ket{i}}{E_{n}-E_{i}-\omega+i\Gamma},
\end{align}
where $\ket{i}$ is the initial (ground) state of the system, $\ket{f}$ are the possible final states and $\ket{n}$ are intermediate states in the presence of the core hole.
$\mathbf{R}_m$ is the lattice vector for site $m$ and the factor $\chi_{\rho\sigma}$ depends on the specific experimental set up, which separates the signal into spin-flip (SF) and non spin-flip (NSF) channels.
$\Gamma$ is the inverse of core hole lifetime: it represents decay channels that are only taken into account phenomenologically, such as decay through phonon emission. In this paper we take the value $\Gamma\sim 0.2\text{eV}$.

Following Ref.~\onlinecite{Benjamin2014}, the intensity can be expressed as an integral:
\begin{align}
\label{eq:intensity-time}
I (\bold{Q},\Delta\omega)\propto& \int_{-\infty}^\infty \!\!\!\!\text{d}s \int_{0}^\infty \!\!\!\text{d}t \int_{0}^\infty \!\!\!\text{d}\tau 
e^{\text{i}\omega (t-\tau) - \text{i}s\Delta\omega - \Gamma(t+\tau)} \nonumber\\
&\times \sum_{m,n}\chi_{\rho\sigma}\chi_{\mu\nu} e^{\text{i}\bQ\cdot(\mathbf{R}_m-\mathbf{R}_n)} S^{mn}_{\rho\sigma\mu\nu},
\end{align}
where $S^{mn}_{\rho\sigma\mu\nu}$ involves evolution of the system before, during and after the absorption of the X-ray and the excitation of the core hole (for details see Ref. \onlinecite{Benjamin2014}). 
%{\color{red}In the specific case of the YRZ propagator, we work in the approximation that the bosonic holons are condensed and that the local variation of this condensate due to the presence of the core hole is incorporated in a renormalization of the core hole potential $U_c$,} 
\begin{align}\label{Smn}
&S^{mn}_{\rho\sigma\mu\nu}\sim\\ \nonumber
& g^2_t(x)\langle e^{\text{i}H\tau} c_{n\rho} e^{-\text{i}H_n\tau} c_{n\sigma}^\dagger e^{\text{i}Hs} c_{m\mu} e^{\text{i}H_mt} c_{m\nu}^\dagger e^{-\text{i}H(t+s)}\rangle.
\end{align}
Here $H_{m(n)}$ is the intermediate hamiltonian in the presence of a core hole at site $m(n)$.
Usually it is assumed that the core hole provides an attractive point potential: $H_m=H_{cd} + \sum_\sigma U_c c^\dagger_{m\sigma}c_{m\sigma}$ (with $U_c<0$).
In this work, there is also the possibility that the core hole leads to a potential for the auxiliary fermions, 
$H_m=H_{cd} + \sum_\sigma U_c c^\dagger_{m\sigma}c_{m\sigma}+\sum_\sigma U_d d^\dagger_{m\sigma}d_{m\sigma}$.
However in the absence of a strong physical motivation we neglect such an effect.

We can now evaluate Eq. \eqref{Smn} numerically, by relating $S^{mn}_{\rho\sigma\mu\nu}$ to determinants and inverses of single particle evolution operators, as detailed in Ref. \onlinecite{Benjamin2014}.
The dimension of these matrices depends linearly on the number of sites in the system, which makes the computation accessible numerically.
Note that the procedure can also be carried out when superconducting pairing terms are included directly in \eqref{Ham} as shown in Ref. \onlinecite{shi2016superconducting}.
\begin{figure} 
\includegraphics[width=0.3\textwidth]{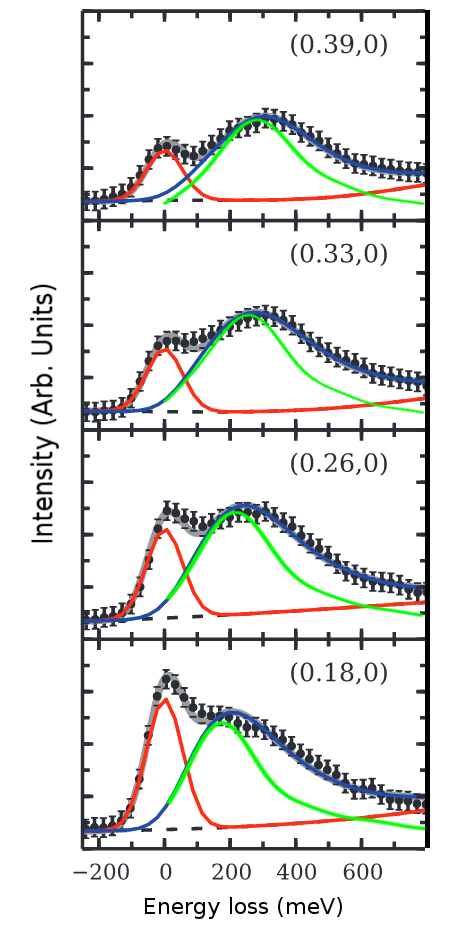}
\caption{RIXS intensity along the $(\zeta,0)$ momentum transfer direction.
The theoretical calculation (green curves) with $U_c=-3\text{eV}$, is compared to experimental data from Ref.~\onlinecite{Dean2014layer} for underdoped Bi-2201 in the pseudogap region (the blue curves are anti-symmetrized Lorentzian fits and the red lines are elastic peaks).}
\label{fig:compare}
\end{figure}

To end this section we calculate the expected RIXS intensity with $U_c=0$, (no core hole potential).
We first solve the hamiltonian in Eq.~\eqref{Ham}, using a linear transformation to new fermionic quasiparticles annihilated by $\alpha_{\bk\sigma}$ and $\beta_{\bk\sigma}$:
\begin{align}
c_{\bk\sigma}&=\cos\theta_{\bk}~\alpha_{\bk\sigma}+\sin\theta_{\bk}~\beta_{\bk\sigma},\nonumber \\
d_{\bk\sigma}&=-\sin\theta_{\bk}~\alpha_{\bk\sigma}+\cos\theta_{\bk}~\beta_{\bk\sigma},
\end{align}
where $\tan 2\theta_{\bk}=\frac{{2\Delta(\bk)}}{2\xi_{0}(\bk)+\xi^{'}(\bk)}$.
The effective hamiltonian is then just:
\begin{align}
\label{eq:Hamdiagonal}
H_{\alpha\beta} = \sum_{\bk \sigma}\epsilon_{+}(\bk)\alpha_{\bk \sigma}^{\dagger}\alpha_{\bk \sigma}+\epsilon_{-}(\bk)\beta_{\bk \sigma}^{\dagger}\beta_{\bk \sigma},
\end{align}
and the energy eigenvalues are:
\begin{align}
\epsilon_{\pm}(\bk)=\frac{{\xi'(\bk)}}{2}\pm\sqrt{\left(\frac{{2\xi_{0}(\bk)+\xi'(\bk)}}{2}\right)^{2}+|\Delta(\bk)|^{2}}.
\end{align}
With these definitions, the scattering amplitude, $A_f$, in Eq.~\eqref{K-H} can be written in terms of $\alpha_{\bk\sigma}$ and $\beta_{\bk\sigma}$, which are the true excitations of the model:
%\begin{align}\label{Lowest}
%A_{f}=&\sum_{\bk}\bra{f}(\text{{cos}}\theta_{\bk}\alpha_{\bk\rho}+\text{{sin}}\theta_{\bk}\beta_{\bk\rho})\nonumber\\
%&\times\sum_{n}\frac{{\ket{n}\bra{n}}}{E_{n}-E_{i}-\omega_i+i\Gamma}\nonumber\\
%&\times(\text{{sin}}\theta_{\bk+\bQ}\alpha_{\bk+\bQ\sigma}^{\dagger}+\text{{cos}}\theta_{\bk+\bQ}\beta_{\bk+\bQ\sigma}^{\dagger})| i\rangle.
%\end{align}
\begin{align}\label{Lowest}
A_{f}=&\sum_{\bk}\chi_{\rho \sigma}\bra{f}(\cos\theta_{\bk+\bQ}\alpha_{\bk+\bQ\rho}+\sin\theta_{\bk+\bQ}\beta_{\bk+\bQ\rho})\nonumber\\
&\times\sum_{n}\frac{\ket{n}\bra{n}}{E_{n}-E_{i}-\omega+i\Gamma}\nonumber\\
&\times(\cos\theta_{\bk}\alpha_{\bk\sigma}^{\dagger}+\sin\theta_{\bk}\beta_{\bk\sigma}^{\dagger})| i\rangle.
\end{align}
To evaluate this expression we take advantage of the fact that for $U_c=0$ we can choose the intermediate states, $\ket{n}$ to be eigenstates of Eq.~\eqref{eq:Hamdiagonal}.
Furthermore, specialising to the case of the YRZ model in the underdoped regime, only the lower $\epsilon_-(\bk)$ band of the initial state $\ket{i}$ is occupied at $T=0$, so we only need to retain terms that include a $\beta$ annihilation operator, as these represent transitions that originate in the $\epsilon_-$ band.
We also neglect a contribution proportional to $\delta_{\bQ,0}$.
The result is
\begin{align}\label{eq:zerohole}
A_{f}=\sum_{\bk}\chi_{\rho \sigma}&\Big\{\frac{\bra{f} \sin \theta_{\bk+\bQ} \cos \theta_{\bk} \beta_{\bk+\bQ\rho} \alpha^\dagger_{\bk\sigma}\ket{i}}{\omega-i\Gamma-\epsilon_+(\bk)}\nonumber\\
& + \frac{\bra{f} \sin \theta_{\bk+\bQ} \sin \theta_{\bk}\beta_{\bk+\bQ\rho} \beta^\dagger_{\bk\sigma}\ket{i}}{\omega-i\Gamma-\epsilon_-(\bk)} \Big\},
\end{align}
The total RIXS intensity is then the summation of $\absval{A_{f}}^2$ over all possible final states $\ket{f}$, including conservation of energy:
\begin{align}
\label{eq:zh-intensity}
I(\bQ,\Delta\omega)& \propto \sum_\bk \absval{\chi_{\sigma\rho}}^2 \Big\{\nonumber\\
&\delta\big(\epsilon_+(\bk)-\epsilon_-(\bk+\bQ)-\Delta\omega\big)\nonumber\\
\times&\frac{\Theta\big(\epsilon_+(\bk)\big)\Theta\big(-\epsilon_-(\bk+\bQ)\big)\sin^2\theta_{\bk+\bQ}\cos^2\theta_{\bk}}{\big(\omega-\epsilon_+(\bk)\big)^2+\Gamma^2}\nonumber\\
+&\delta\big(\epsilon_-(\bk)-\epsilon_-(\bk+\bQ)-\Delta\omega\big)\nonumber\\
\times&\frac{\Theta\big(\epsilon_-(\bk)\big)\Theta\big(-\epsilon_-(\bk+\bQ)\big)\sin^2\theta_{\bk+\bQ}\sin^2\theta_{\bk}}{\big(\omega-\epsilon_-(\bk)\big)^2+\Gamma^2}\Big\}.
\end{align}
The extension to the more general case including transitions starting in the $\epsilon_+$ band is simple.
In practice the delta functions are replaced by Gaussians to reflect an experimental resolution of 150meV.

In the limit that $\Gamma \gg \omega, \epsilon_{\pm}$ Eq.~\eqref{eq:zh-intensity} gives essentially the same result as the calculation of the YRZ spin dynamical structure factor in Refs.~\onlinecite{James2012} and \onlinecite{DeanBi2212}, except for an effective `random phase approximation' (RPA) resummation of the susceptibility that occurs in those works (see also Ref.~\onlinecite{brinckmann2001}).

\section{Comparison with RIXS data}
We evaluate Eq.~\eqref{eq:intensity-time} numerically, using the determinant method described in Ref.~\onlinecite{Benjamin2014}.
Fig.~\ref{fig:compare} shows a comparison between this calculation and experimental data for hole doped Bi-2201 (at underdoping $x=0.12$) reported in Ref.~\onlinecite{Dean2014layer}.
Quantitative agreement with the experiments was reported using the itinerant quasiparticle approach in Ref.~\onlinecite{Benjamin2014} and with a calculation of the YRZ dynamical spin susceptibility in Ref.~\onlinecite{Dean2014layer}.
Next, we show how the combined approach improves on the YRZ-based result.
We emphasize, though, that our calculation relies on the YRZ parameters used in Ref.~\onlinecite{Dean2014layer} (and originally taken from fits to ARPES data\cite{Yang2008}), and are essentially those pertaining to Bi-2212 bilayers, while the experiments have been carried out on Bi-2201.
Better determined tight-binding parameters would be essential for a real test of the YRZ approach, but are outside the scope of this paper.

\begin{figure} 
\includegraphics[width=0.4\textwidth]{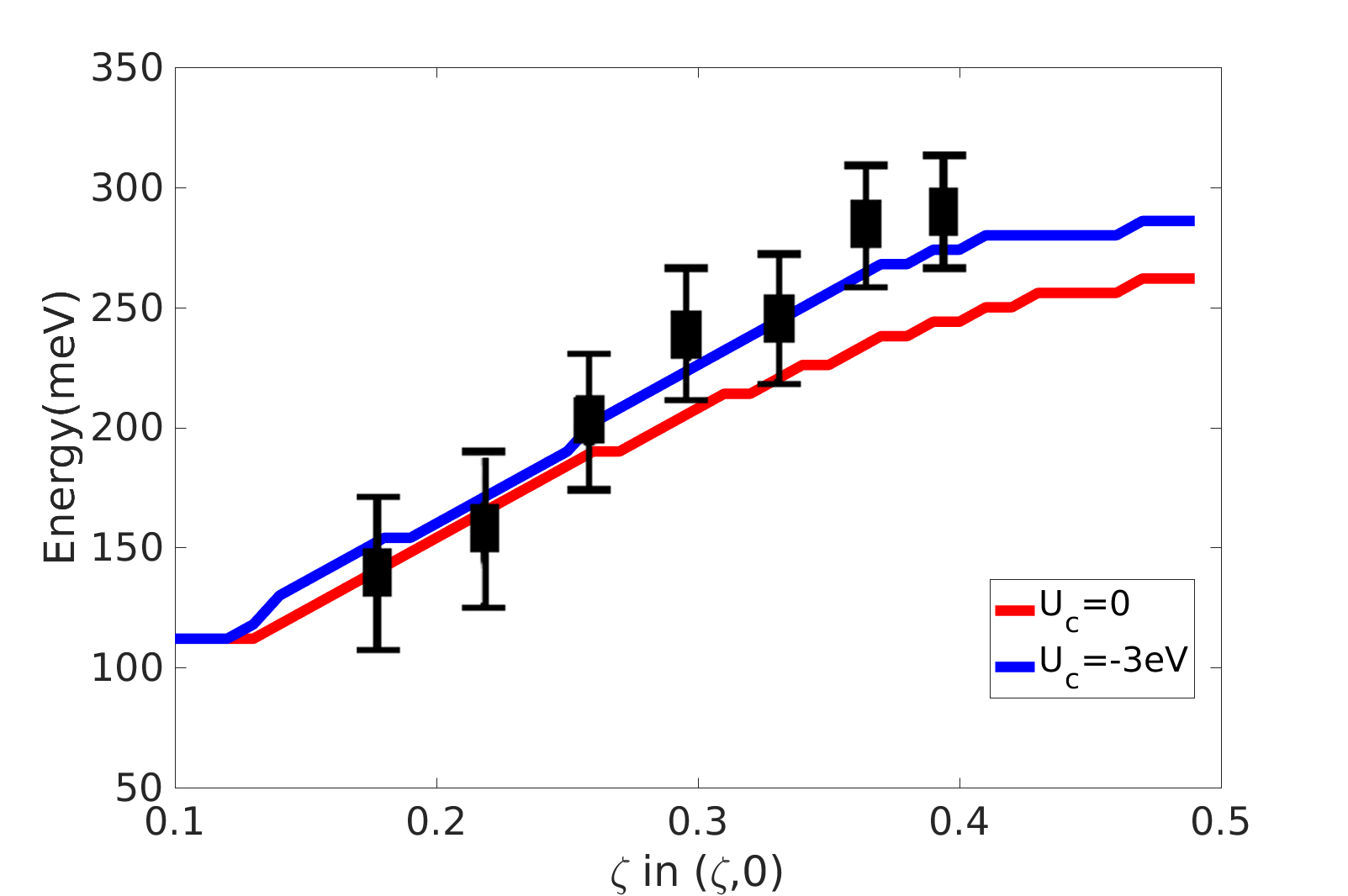}
\caption{The dispersion of the paramagnon mode in underdoped Bi-2201, along $(\zeta,0)$.
The red line shows the calculation with no core hole potential, (giving a similar result to Ref.~\onlinecite{Dean2014layer}).
The blue line shows the peak position for the spin-flip contribution with a core hole potential $U_c=-3\text{eV}$. Experimental data reported in Ref.~\onlinecite{Dean2014layer} are noted by black squares.}
\label{fig:corehole}
\end{figure}
\begin{figure}[tb]
\includegraphics[width=0.43\textwidth]{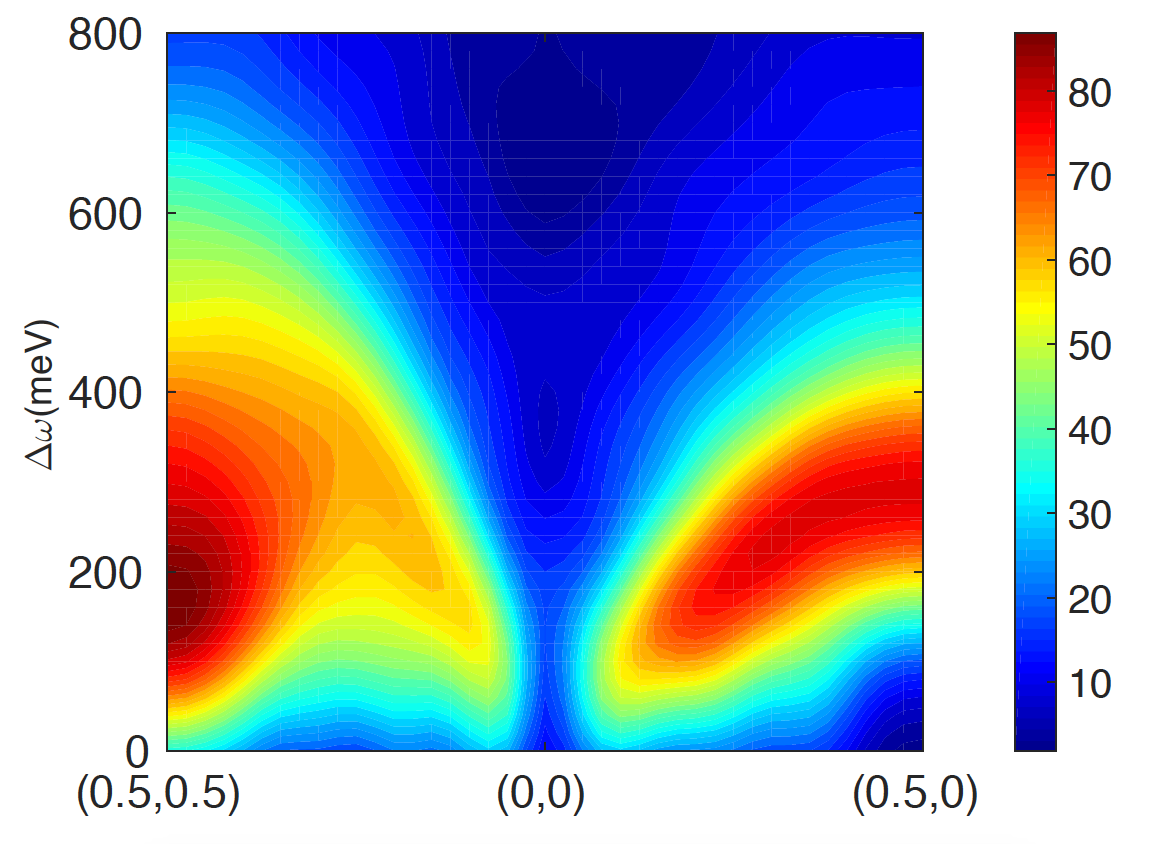}
\caption{RIXS intensity along the $(1,1)$ and $(1,0)$ directions, calculated using the parameters in Table \ref{param} and $U_c=0$ (effectively indistinguishable from the $U_c=-3\text{eV}$ case when presented as a color density plot).
Although there is a clear peak along the antinodal $(1,0)$ direction, it is difficult to interpret the intensity along the (1,1) direction in terms of damped spin wave excitations (finite lifetime magnons).}
\label{fig:2d}
\end{figure}

To argue the necessity of taking into account the core hole dynamics, we show in Fig.~\ref{fig:corehole} the effect of adding a core hole, by comparing the intensities calculated using Eqs.~\eqref{eq:intensity-time} and \eqref{Smn} versus Eq.~\eqref{eq:zh-intensity}.
As we remarked on in the previous section, the latter gives a result similar to previous YRZ spin susceptibility calculations, except for an effective `random phase approximation' (RPA) resummation\cite{James2012}.
However for cuprate parameters the effect of this RPA is confined to low energies\cite{DeanBi2212} $< 100 \text{meV}$ and therefore it is not a significant factor in our comparison.
We find that the core hole pushes the peaks to higher energy transfer and that this effect is more significant at large momentum transfer.
While the calculation with $U_c=0$ catches the essential response, the inclusion of a core hole significantly improves the agreement with the experiment at high energies.

The shift of the response to higher energies can be understood by expanding Eq.~\eqref{K-H} to first order in the core hole potential $V_r=U_c \sum_\sigma c_{r\sigma}^{\dagger}c_{r\sigma}$.
This yields a term:
\begin{align}
A_f^1 \propto &U_c \sum_{m}\chi_{\rho\sigma} e^{i\bQ\cdot\mathbf{R}_m}\nonumber\\
& \times\bra{f}c_{m\rho}G_0(c_{m\uparrow}^{\dagger}c_{m\uparrow}+c_{m\downarrow}^{\dagger}c_{m\downarrow})G_0c_{m\sigma}^{\dagger}\ket{i},
\end{align}
where $G_0^{-1}=\omega-i\Gamma-(H_{cd}-E_i)$.
This means the final state would have two pairs of quasi particle-hole excitations, with total momentum $\bold{Q}$ and total energy $\Delta\omega$, while in the no core hole case, the excitations are a single quasi particle-hole pair, with the same total energy and momentum, and the excitations are mostly close to the Fermi surface. The core hole allows the individual excitations to explore a larger phase space, further from the Fermi surface, and thus the excitation energies are higher, and the peak moves to the right. This effect is much harder to analyze quantitatively, but the determinant method allows us to calculate it numerically.

It is important to understand the relation between the present treatment and other RIXS calculations, based on magnetic susceptibility, such as carried out in e.g. Ref.~\onlinecite{James2012}. We point out that in the case where $U_c = 0$, i.e.  no core hole, our approach yields a result that is similar to the dynamical susceptibility: In Eq.~\eqref{K-H}, if we assume $\Gamma$ is much larger than the other energy scales of the system, we can replace the term $(E_{n}-E_{i}-\omega+i\Gamma)^{-1}$ by $F(\omega,\Gamma)$ for any $\ket{n}$, and the intensity is written as the Fourier transform of the 4-point function:
\bea\label{nocorehole}
I(\mathbf{R}_{nm},t)=F(\omega,\Gamma)\chi_{\sigma\lambda}\chi_{\mu\nu}\langle\rho_{n\sigma\lambda}(t)\rho_{m\mu\nu}(0)\rangle,
\eea
where $\rho_{n\tau\sigma}=c_{n\tau}^{\dagger}c_{n\sigma}$, and we have used that $\delta(E)=\int_{-\infty}^{\infty}  {\text{d}t\over 2\pi}    e^{-iEt}$.
In Ref.~\onlinecite{James2012} the irreducible part of the magnetic susceptibility is defined as:
\begin{align}\label{suscept}
\chi^{\text{irr}}\!(\mathbf{R}_{nm},t)=i\langle T(\rho_{n\uparrow\uparrow}(t)\!-\!\rho_{n\downarrow\downarrow}(t)) (\rho_{m\uparrow\uparrow}(0)\!-\!\rho_{m\downarrow\downarrow}(0))\rangle,
\end{align}
where $T$ indicates time ordering.
Eqs. \eqref{nocorehole} and \eqref{suscept} are both density-density correlation functions of the system, and have similar behavior.

In Fig.~\ref{fig:2d}, we show the calculated intensity along high symmetry lines for $U_c=0$ (a color density plot of the $U_c=-3\text{eV}$ case would be indistinguishable).
Similar to the conclusions in Refs.~\onlinecite{Dean2014layer} and \onlinecite{Guarise2014cuprate}, we see that along the nodal $(1,1)$ direction the RIXS spectrum becomes much more diffuse and less sensitive to momentum transfer, which is difficult to understand in the framework of finite lifetime broadened magnon excitations (damped spin waves).
While the RIXS signal is commonly interpreted as a primarily magnetic response\cite{Le-Tacon:2011aa,dean2013persistence,Ellis2015}, here we see that our tight binding hamiltonian approach can quantitatively explore the RIXS spectrum for various momentum transfers, and go beyond simple spin wave theories.

\section{Conclusions}
We have shown how to calculate the RIXS response, including a dynamical treatment of the transient core hole, of systems with non-trivial single particle Green's functions that feature both zeros and poles.
We do so by introducing auxiliary fermions, yielding a tight-binding formulation that can be treated by the method of Ref.~\onlinecite{Benjamin2014}.
Our approach is appropriate to systems and models where many-body interaction effects can primarily be described through renormalization of the single particle propagator (\emph{i.e.} through dressed quasiparticles), and does not incorporate higher order (quasi) particle hole `bubble' diagrams.

As a test of our approach we applied it to the YRZ ansatz, a semi-phenomenological Green's function popular in studies of high-$T_c$ cuprates.
Comparing our results to experiments on Bi-2201, we showed that inclusion of the core hole potential moves dispersion peaks to higher energy along the $(1,0)$ momentum transfer direction, giving better agreement with the experimental data than previous calculations based on YRZ physics.
Examining the effect of the core hole potential perturbatively, we see that this shift to higher energy is due to an enhancement of the scattering phase space, suggesting that it might be a general feature in calculated RIXS spectra.

Finally, we note that this method can also be used to describe systems in which the auxiliary fermions have a definite physical manifestation, for example coupled systems where the RIXS probe only interacts with one species (band) of fermions.

%{\color{green}We have studied the non-equilibrium dynamics associated with the YRZ ansatz in the presence of X-ray absorption by introducing phenomenological tight-binding hamiltonian model involving auxiliary fermions. This approach allows us to resolve the non-locality of the action in time and is particularly useful when dealing with the core hole introduced in RIXS experiments. We compare a theoretical computation based on our model with experiments on Bi-2201, and show that the core hole moves dispersion peaks to higher energy in $(1,0)$ direction, giving a better agreement with the experimental data. In addition we observe that in the $(1,1)$ direction the signal is more diffused and a well-defined magnon peak is absent.}

Acknowledgements: We have benefitted from discussions with R. M. Konik, M. P. M. Dean and M. Kan\'{a}sz-Nagy.
The work of IK and YS was supported by the NSF grants DMR-1508245 and CAREER DMR-0956053.
AJAJ was supported by the UK Engineering and Physical Sciences Research Council, fellowship no. EP/L010623/1. ED acknowledges support from the
Harvard-MIT CUA, NSF Grant No. DMR-1308435, and the AFOSR Quantum Simulation MURI.

\bibliography{YRZ_AJAJ}

\begin{thebibliography}{29}
\expandafter\ifx\csname natexlab\endcsname\relax\def\natexlab#1{#1}\fi
\expandafter\ifx\csname bibnamefont\endcsname\relax
  \def\bibnamefont#1{#1}\fi
\expandafter\ifx\csname bibfnamefont\endcsname\relax
  \def\bibfnamefont#1{#1}\fi
\expandafter\ifx\csname citenamefont\endcsname\relax
  \def\citenamefont#1{#1}\fi
\expandafter\ifx\csname url\endcsname\relax
  \def\url#1{\texttt{#1}}\fi
\expandafter\ifx\csname urlprefix\endcsname\relax\def\urlprefix{URL }\fi
\providecommand{\bibinfo}[2]{#2}
\providecommand{\eprint}[2][]{\url{#2}}

\bibitem[{\citenamefont{Benjamin et~al.}(2014)\citenamefont{Benjamin, Klich,
  and Demler}}]{Benjamin2014}
\bibinfo{author}{\bibfnamefont{D.}~\bibnamefont{Benjamin}},
  \bibinfo{author}{\bibfnamefont{I.}~\bibnamefont{Klich}}, \bibnamefont{and}
  \bibinfo{author}{\bibfnamefont{E.}~\bibnamefont{Demler}},
  \bibinfo{journal}{Phys. Rev. Lett.} \textbf{\bibinfo{volume}{112}},
  \bibinfo{pages}{247002} (\bibinfo{year}{2014}).

\bibitem[{\citenamefont{Lake et~al.}(2013)\citenamefont{Lake, Tennant, Caux,
  Barthel, Schollw{\"o}ck, Nagler, and Frost}}]{lake2013multispinon}
\bibinfo{author}{\bibfnamefont{B.}~\bibnamefont{Lake}},
  \bibinfo{author}{\bibfnamefont{D.}~\bibnamefont{Tennant}},
  \bibinfo{author}{\bibfnamefont{J.-S.} \bibnamefont{Caux}},
  \bibinfo{author}{\bibfnamefont{T.}~\bibnamefont{Barthel}},
  \bibinfo{author}{\bibfnamefont{U.}~\bibnamefont{Schollw{\"o}ck}},
  \bibinfo{author}{\bibfnamefont{S.}~\bibnamefont{Nagler}}, \bibnamefont{and}
  \bibinfo{author}{\bibfnamefont{C.}~\bibnamefont{Frost}},
  \bibinfo{journal}{Phys. Rev. Lett.} \textbf{\bibinfo{volume}{111}},
  \bibinfo{pages}{137205} (\bibinfo{year}{2013}).

\bibitem[{\citenamefont{Dean}(2015)}]{dean2015insights}
\bibinfo{author}{\bibfnamefont{M.}~\bibnamefont{Dean}}, \bibinfo{journal}{J.
  Magn. Magn. Mater.} \textbf{\bibinfo{volume}{376}}, \bibinfo{pages}{3}
  (\bibinfo{year}{2015}).

\bibitem[{\citenamefont{Vernay et~al.}(2008)\citenamefont{Vernay, Moritz,
  Elfimov, Geck, Hawthorn, Devereaux, and Sawatzky}}]{vernay2008cu}
\bibinfo{author}{\bibfnamefont{F.}~\bibnamefont{Vernay}},
  \bibinfo{author}{\bibfnamefont{B.}~\bibnamefont{Moritz}},
  \bibinfo{author}{\bibfnamefont{I.}~\bibnamefont{Elfimov}},
  \bibinfo{author}{\bibfnamefont{J.}~\bibnamefont{Geck}},
  \bibinfo{author}{\bibfnamefont{D.}~\bibnamefont{Hawthorn}},
  \bibinfo{author}{\bibfnamefont{T.}~\bibnamefont{Devereaux}},
  \bibnamefont{and} \bibinfo{author}{\bibfnamefont{G.}~\bibnamefont{Sawatzky}},
  \bibinfo{journal}{Phys. Rev. B} \textbf{\bibinfo{volume}{77}},
  \bibinfo{pages}{104519} (\bibinfo{year}{2008}).

\bibitem[{\citenamefont{Chen et~al.}(2010)\citenamefont{Chen, Moritz, Vernay,
  Hancock, Johnston, Jia, Chabot-Couture, Greven, Elfimov, Sawatzky
  et~al.}}]{chen2010unraveling}
\bibinfo{author}{\bibfnamefont{C.-C.} \bibnamefont{Chen}},
  \bibinfo{author}{\bibfnamefont{B.}~\bibnamefont{Moritz}},
  \bibinfo{author}{\bibfnamefont{F.}~\bibnamefont{Vernay}},
  \bibinfo{author}{\bibfnamefont{J.~N.} \bibnamefont{Hancock}},
  \bibinfo{author}{\bibfnamefont{S.}~\bibnamefont{Johnston}},
  \bibinfo{author}{\bibfnamefont{C.~J.} \bibnamefont{Jia}},
  \bibinfo{author}{\bibfnamefont{G.}~\bibnamefont{Chabot-Couture}},
  \bibinfo{author}{\bibfnamefont{M.}~\bibnamefont{Greven}},
  \bibinfo{author}{\bibfnamefont{I.}~\bibnamefont{Elfimov}},
  \bibinfo{author}{\bibfnamefont{G.~A.} \bibnamefont{Sawatzky}},
  \bibnamefont{et~al.}, \bibinfo{journal}{Phys. Rev. Lett.}
  \textbf{\bibinfo{volume}{105}}, \bibinfo{pages}{177401}
  (\bibinfo{year}{2010}).

\bibitem[{\citenamefont{Guarise et~al.}(2014)\citenamefont{Guarise,
  Dalla~Piazza, Berger, Giannini, Schmitt, R{\o}nnow, Sawatzky, van~den Brink,
  Altenfeld, Eremin et~al.}}]{Guarise2014cuprate}
\bibinfo{author}{\bibfnamefont{M.}~\bibnamefont{Guarise}},
  \bibinfo{author}{\bibfnamefont{B.}~\bibnamefont{Dalla~Piazza}},
  \bibinfo{author}{\bibfnamefont{H.}~\bibnamefont{Berger}},
  \bibinfo{author}{\bibfnamefont{E.}~\bibnamefont{Giannini}},
  \bibinfo{author}{\bibfnamefont{T.}~\bibnamefont{Schmitt}},
  \bibinfo{author}{\bibfnamefont{H.}~\bibnamefont{R{\o}nnow}},
  \bibinfo{author}{\bibfnamefont{G.}~\bibnamefont{Sawatzky}},
  \bibinfo{author}{\bibfnamefont{J.}~\bibnamefont{van~den Brink}},
  \bibinfo{author}{\bibfnamefont{D.}~\bibnamefont{Altenfeld}},
  \bibinfo{author}{\bibfnamefont{I.}~\bibnamefont{Eremin}},
  \bibnamefont{et~al.}, \bibinfo{journal}{Nature Communications}
  \textbf{\bibinfo{volume}{5}} (\bibinfo{year}{2014}).

\bibitem[{\citenamefont{Ament et~al.}(2011)\citenamefont{Ament, van Veenendaal,
  Devereaux, Hill, and van~den Brink}}]{ament2011resonant}
\bibinfo{author}{\bibfnamefont{L.~J.} \bibnamefont{Ament}},
  \bibinfo{author}{\bibfnamefont{M.}~\bibnamefont{van Veenendaal}},
  \bibinfo{author}{\bibfnamefont{T.~P.} \bibnamefont{Devereaux}},
  \bibinfo{author}{\bibfnamefont{J.~P.} \bibnamefont{Hill}}, \bibnamefont{and}
  \bibinfo{author}{\bibfnamefont{J.}~\bibnamefont{van~den Brink}},
  \bibinfo{journal}{Reviews of Modern Physics} \textbf{\bibinfo{volume}{83}},
  \bibinfo{pages}{705} (\bibinfo{year}{2011}).

\bibitem[{\citenamefont{Kan{\'a}sz-Nagy
  et~al.}(2016)\citenamefont{Kan{\'a}sz-Nagy, Shi, Klich, and
  Demler}}]{kanasz2016resonant}
\bibinfo{author}{\bibfnamefont{M.}~\bibnamefont{Kan{\'a}sz-Nagy}},
  \bibinfo{author}{\bibfnamefont{Y.}~\bibnamefont{Shi}},
  \bibinfo{author}{\bibfnamefont{I.}~\bibnamefont{Klich}}, \bibnamefont{and}
  \bibinfo{author}{\bibfnamefont{E.}~\bibnamefont{Demler}},
  \bibinfo{journal}{Phys. Rev. B} \textbf{\bibinfo{volume}{94}},
  \bibinfo{pages}{165127} (\bibinfo{year}{2016}).

\bibitem[{\citenamefont{Jia et~al.}(2016)\citenamefont{Jia, Wohlfeld, Wang,
  Moritz, and Devereaux}}]{jia2016using}
\bibinfo{author}{\bibfnamefont{C.}~\bibnamefont{Jia}},
  \bibinfo{author}{\bibfnamefont{K.}~\bibnamefont{Wohlfeld}},
  \bibinfo{author}{\bibfnamefont{Y.}~\bibnamefont{Wang}},
  \bibinfo{author}{\bibfnamefont{B.}~\bibnamefont{Moritz}}, \bibnamefont{and}
  \bibinfo{author}{\bibfnamefont{T.~P.} \bibnamefont{Devereaux}},
  \bibinfo{journal}{Phys. Rev. X} \textbf{\bibinfo{volume}{6}},
  \bibinfo{pages}{021020} (\bibinfo{year}{2016}).

\bibitem[{\citenamefont{Shi et~al.}(2016)\citenamefont{Shi, Benjamin, Demler,
  and Klich}}]{shi2016superconducting}
\bibinfo{author}{\bibfnamefont{Y.}~\bibnamefont{Shi}},
  \bibinfo{author}{\bibfnamefont{D.}~\bibnamefont{Benjamin}},
  \bibinfo{author}{\bibfnamefont{E.}~\bibnamefont{Demler}}, \bibnamefont{and}
  \bibinfo{author}{\bibfnamefont{I.}~\bibnamefont{Klich}},
  \bibinfo{journal}{Phys. Rev. B} \textbf{\bibinfo{volume}{94}},
  \bibinfo{pages}{094516} (\bibinfo{year}{2016}).

\bibitem[{\citenamefont{Yang et~al.}(2006)\citenamefont{Yang, Rice, and
  Zhang}}]{YRZ}
\bibinfo{author}{\bibfnamefont{K.-Y.} \bibnamefont{Yang}},
  \bibinfo{author}{\bibfnamefont{T.}~\bibnamefont{Rice}}, \bibnamefont{and}
  \bibinfo{author}{\bibfnamefont{F.-C.} \bibnamefont{Zhang}},
  \bibinfo{journal}{Phys. Rev. B} \textbf{\bibinfo{volume}{73}},
  \bibinfo{pages}{174501} (\bibinfo{year}{2006}).

\bibitem[{\citenamefont{Norman et~al.}(1998)\citenamefont{Norman, Ding,
  Randeria, Campuzano, Yokoya, Takeuchi, Takahashi, Mochiku, Kadowaki,
  Guptasarma et~al.}}]{norman1998destruction}
\bibinfo{author}{\bibfnamefont{M.}~\bibnamefont{Norman}},
  \bibinfo{author}{\bibfnamefont{H.}~\bibnamefont{Ding}},
  \bibinfo{author}{\bibfnamefont{M.}~\bibnamefont{Randeria}},
  \bibinfo{author}{\bibfnamefont{J.}~\bibnamefont{Campuzano}},
  \bibinfo{author}{\bibfnamefont{T.}~\bibnamefont{Yokoya}},
  \bibinfo{author}{\bibfnamefont{T.}~\bibnamefont{Takeuchi}},
  \bibinfo{author}{\bibfnamefont{T.}~\bibnamefont{Takahashi}},
  \bibinfo{author}{\bibfnamefont{T.}~\bibnamefont{Mochiku}},
  \bibinfo{author}{\bibfnamefont{K.}~\bibnamefont{Kadowaki}},
  \bibinfo{author}{\bibfnamefont{P.}~\bibnamefont{Guptasarma}},
  \bibnamefont{et~al.}, \bibinfo{journal}{Nature}
  \textbf{\bibinfo{volume}{392}}, \bibinfo{pages}{157} (\bibinfo{year}{1998}).

\bibitem[{\citenamefont{Konik et~al.}(2006)\citenamefont{Konik, Rice, and
  Tsvelik}}]{KRT2006}
\bibinfo{author}{\bibfnamefont{R.}~\bibnamefont{Konik}},
  \bibinfo{author}{\bibfnamefont{T.}~\bibnamefont{Rice}}, \bibnamefont{and}
  \bibinfo{author}{\bibfnamefont{A.}~\bibnamefont{Tsvelik}},
  \bibinfo{journal}{Phys. Rev. Lett.} \textbf{\bibinfo{volume}{96}},
  \bibinfo{pages}{086407} (\bibinfo{year}{2006}).

\bibitem[{\citenamefont{Yang et~al.}(2008)\citenamefont{Yang, Rameau, Johnson,
  Valla, Tsvelik, and Gu}}]{Yang2008}
\bibinfo{author}{\bibfnamefont{H.-B.} \bibnamefont{Yang}},
  \bibinfo{author}{\bibfnamefont{J.}~\bibnamefont{Rameau}},
  \bibinfo{author}{\bibfnamefont{P.}~\bibnamefont{Johnson}},
  \bibinfo{author}{\bibfnamefont{T.}~\bibnamefont{Valla}},
  \bibinfo{author}{\bibfnamefont{A.}~\bibnamefont{Tsvelik}}, \bibnamefont{and}
  \bibinfo{author}{\bibfnamefont{G.}~\bibnamefont{Gu}},
  \bibinfo{journal}{Nature} \textbf{\bibinfo{volume}{456}}, \bibinfo{pages}{77}
  (\bibinfo{year}{2008}).

\bibitem[{\citenamefont{Rice et~al.}(2011)\citenamefont{Rice, Yang, and
  Zhang}}]{rice2011phenomenological}
\bibinfo{author}{\bibfnamefont{T.}~\bibnamefont{Rice}},
  \bibinfo{author}{\bibfnamefont{K.-Y.} \bibnamefont{Yang}}, \bibnamefont{and}
  \bibinfo{author}{\bibfnamefont{F.-C.} \bibnamefont{Zhang}},
  \bibinfo{journal}{Rep. Prog. Phys.} \textbf{\bibinfo{volume}{75}},
  \bibinfo{pages}{016502} (\bibinfo{year}{2011}).

\bibitem[{\citenamefont{Miller et~al.}(2017)\citenamefont{Miller, Zhang,
  Eisaki, and Lanzara}}]{miller2017particle}
\bibinfo{author}{\bibfnamefont{T.~L.} \bibnamefont{Miller}},
  \bibinfo{author}{\bibfnamefont{W.}~\bibnamefont{Zhang}},
  \bibinfo{author}{\bibfnamefont{H.}~\bibnamefont{Eisaki}}, \bibnamefont{and}
  \bibinfo{author}{\bibfnamefont{A.}~\bibnamefont{Lanzara}},
  \bibinfo{journal}{Phys. Rev. Lett.} \textbf{\bibinfo{volume}{118}},
  \bibinfo{pages}{097001} (\bibinfo{year}{2017}).

\bibitem[{\citenamefont{Vishik et~al.}(2012)\citenamefont{Vishik, Hashimoto,
  He, Lee, Schmitt, Lu, Moore, Zhang, Meevasana, Sasagawa
  et~al.}}]{vishik2012phase}
\bibinfo{author}{\bibfnamefont{I.}~\bibnamefont{Vishik}},
  \bibinfo{author}{\bibfnamefont{M.}~\bibnamefont{Hashimoto}},
  \bibinfo{author}{\bibfnamefont{R.-H.} \bibnamefont{He}},
  \bibinfo{author}{\bibfnamefont{W.-S.} \bibnamefont{Lee}},
  \bibinfo{author}{\bibfnamefont{F.}~\bibnamefont{Schmitt}},
  \bibinfo{author}{\bibfnamefont{D.}~\bibnamefont{Lu}},
  \bibinfo{author}{\bibfnamefont{R.}~\bibnamefont{Moore}},
  \bibinfo{author}{\bibfnamefont{C.}~\bibnamefont{Zhang}},
  \bibinfo{author}{\bibfnamefont{W.}~\bibnamefont{Meevasana}},
  \bibinfo{author}{\bibfnamefont{T.}~\bibnamefont{Sasagawa}},
  \bibnamefont{et~al.}, \bibinfo{journal}{P. Natl. Acad. Sci. USA}
  \textbf{\bibinfo{volume}{109}}, \bibinfo{pages}{18332}
  (\bibinfo{year}{2012}).

\bibitem[{\citenamefont{Reber et~al.}(2012)\citenamefont{Reber, Plumb, Sun,
  Cao, Wang, McElroy, Iwasawa, Arita, Wen, Xu et~al.}}]{reber2012origin}
\bibinfo{author}{\bibfnamefont{T.}~\bibnamefont{Reber}},
  \bibinfo{author}{\bibfnamefont{N.}~\bibnamefont{Plumb}},
  \bibinfo{author}{\bibfnamefont{Z.}~\bibnamefont{Sun}},
  \bibinfo{author}{\bibfnamefont{Y.}~\bibnamefont{Cao}},
  \bibinfo{author}{\bibfnamefont{Q.}~\bibnamefont{Wang}},
  \bibinfo{author}{\bibfnamefont{K.}~\bibnamefont{McElroy}},
  \bibinfo{author}{\bibfnamefont{H.}~\bibnamefont{Iwasawa}},
  \bibinfo{author}{\bibfnamefont{M.}~\bibnamefont{Arita}},
  \bibinfo{author}{\bibfnamefont{J.}~\bibnamefont{Wen}},
  \bibinfo{author}{\bibfnamefont{Z.}~\bibnamefont{Xu}}, \bibnamefont{et~al.},
  \bibinfo{journal}{Nat. Phys.} \textbf{\bibinfo{volume}{8}},
  \bibinfo{pages}{606} (\bibinfo{year}{2012}).

\bibitem[{\citenamefont{Dalla~Torre et~al.}(2016)\citenamefont{Dalla~Torre,
  Benjamin, He, Dentelski, and Demler}}]{dalla2016friedel}
\bibinfo{author}{\bibfnamefont{E.}~\bibnamefont{Dalla~Torre}},
  \bibinfo{author}{\bibfnamefont{D.}~\bibnamefont{Benjamin}},
  \bibinfo{author}{\bibfnamefont{Y.}~\bibnamefont{He}},
  \bibinfo{author}{\bibfnamefont{D.}~\bibnamefont{Dentelski}},
  \bibnamefont{and} \bibinfo{author}{\bibfnamefont{E.}~\bibnamefont{Demler}},
  \bibinfo{journal}{Phys. Rev. B} \textbf{\bibinfo{volume}{93}},
  \bibinfo{pages}{205117} (\bibinfo{year}{2016}).

\bibitem[{\citenamefont{James et~al.}(2012)\citenamefont{James, Konik, and
  Rice}}]{James2012}
\bibinfo{author}{\bibfnamefont{A.}~\bibnamefont{James}},
  \bibinfo{author}{\bibfnamefont{R.}~\bibnamefont{Konik}}, \bibnamefont{and}
  \bibinfo{author}{\bibfnamefont{T.}~\bibnamefont{Rice}},
  \bibinfo{journal}{Phys. Rev. B} \textbf{\bibinfo{volume}{86}},
  \bibinfo{pages}{100508} (\bibinfo{year}{2012}).

\bibitem[{\citenamefont{Brinckmann and Lee}(2001)}]{brinckmann2001}
\bibinfo{author}{\bibfnamefont{J.}~\bibnamefont{Brinckmann}} \bibnamefont{and}
  \bibinfo{author}{\bibfnamefont{P.}~\bibnamefont{Lee}},
  \bibinfo{journal}{Phys. Rev. B} \textbf{\bibinfo{volume}{65}},
  \bibinfo{pages}{014502} (\bibinfo{year}{2001}).

\bibitem[{\citenamefont{Dean et~al.}(2013{\natexlab{a}})\citenamefont{Dean,
  James, Springell, Liu, Monney, Zhou, Konik, Wen, Xu, Gu et~al.}}]{DeanBi2212}
\bibinfo{author}{\bibfnamefont{M.}~\bibnamefont{Dean}},
  \bibinfo{author}{\bibfnamefont{A.}~\bibnamefont{James}},
  \bibinfo{author}{\bibfnamefont{R.}~\bibnamefont{Springell}},
  \bibinfo{author}{\bibfnamefont{X.}~\bibnamefont{Liu}},
  \bibinfo{author}{\bibfnamefont{C.}~\bibnamefont{Monney}},
  \bibinfo{author}{\bibfnamefont{K.}~\bibnamefont{Zhou}},
  \bibinfo{author}{\bibfnamefont{R.}~\bibnamefont{Konik}},
  \bibinfo{author}{\bibfnamefont{J.}~\bibnamefont{Wen}},
  \bibinfo{author}{\bibfnamefont{Z.}~\bibnamefont{Xu}},
  \bibinfo{author}{\bibfnamefont{G.}~\bibnamefont{Gu}}, \bibnamefont{et~al.},
  \bibinfo{journal}{Phys. Rev. Lett.} \textbf{\bibinfo{volume}{110}},
  \bibinfo{pages}{147001} (\bibinfo{year}{2013}{\natexlab{a}}).

\bibitem[{\citenamefont{Dean et~al.}(2014)\citenamefont{Dean, James, Walters,
  Bisogni, Jarrige, H{\"u}cker, Giannini, Fujita, Pelliciari, Huang
  et~al.}}]{Dean2014layer}
\bibinfo{author}{\bibfnamefont{M.}~\bibnamefont{Dean}},
  \bibinfo{author}{\bibfnamefont{A.}~\bibnamefont{James}},
  \bibinfo{author}{\bibfnamefont{A.}~\bibnamefont{Walters}},
  \bibinfo{author}{\bibfnamefont{V.}~\bibnamefont{Bisogni}},
  \bibinfo{author}{\bibfnamefont{I.}~\bibnamefont{Jarrige}},
  \bibinfo{author}{\bibfnamefont{M.}~\bibnamefont{H{\"u}cker}},
  \bibinfo{author}{\bibfnamefont{E.}~\bibnamefont{Giannini}},
  \bibinfo{author}{\bibfnamefont{M.}~\bibnamefont{Fujita}},
  \bibinfo{author}{\bibfnamefont{J.}~\bibnamefont{Pelliciari}},
  \bibinfo{author}{\bibfnamefont{Y.}~\bibnamefont{Huang}},
  \bibnamefont{et~al.}, \bibinfo{journal}{Physical Review B}
  \textbf{\bibinfo{volume}{90}}, \bibinfo{pages}{220506}
  (\bibinfo{year}{2014}).

\bibitem[{\citenamefont{Tsvelik}(2006)}]{Tsvelik2006}
\bibinfo{author}{\bibfnamefont{A.~M.} \bibnamefont{Tsvelik}},
  \emph{\bibinfo{title}{Quantum field theory in condensed matter physics}}
  (\bibinfo{publisher}{Cambridge university press}, \bibinfo{year}{2006}).

\bibitem[{\citenamefont{van~den Brink and van
  Veenendaal}(2005)}]{van2005theory}
\bibinfo{author}{\bibfnamefont{J.}~\bibnamefont{van~den Brink}}
  \bibnamefont{and} \bibinfo{author}{\bibfnamefont{M.}~\bibnamefont{van
  Veenendaal}}, \bibinfo{journal}{J. Phys. Chem. Solids}
  \textbf{\bibinfo{volume}{66}}, \bibinfo{pages}{2145} (\bibinfo{year}{2005}).

\bibitem[{\citenamefont{Sakai et~al.}(2016)\citenamefont{Sakai, Civelli, and
  Imada}}]{sakai2016hidden}
\bibinfo{author}{\bibfnamefont{S.}~\bibnamefont{Sakai}},
  \bibinfo{author}{\bibfnamefont{M.}~\bibnamefont{Civelli}}, \bibnamefont{and}
  \bibinfo{author}{\bibfnamefont{M.}~\bibnamefont{Imada}},
  \bibinfo{journal}{Phys. Rev. Lett.} \textbf{\bibinfo{volume}{116}},
  \bibinfo{pages}{057003} (\bibinfo{year}{2016}).

\bibitem[{\citenamefont{Le~Tacon et~al.}(2011)\citenamefont{Le~Tacon,
  Ghiringhelli, Chaloupka, Sala, Hinkov, Haverkort, Minola, Bakr, Zhou,
  Blanco-Canosa et~al.}}]{Le-Tacon:2011aa}
\bibinfo{author}{\bibfnamefont{M.}~\bibnamefont{Le~Tacon}},
  \bibinfo{author}{\bibfnamefont{G.}~\bibnamefont{Ghiringhelli}},
  \bibinfo{author}{\bibfnamefont{J.}~\bibnamefont{Chaloupka}},
  \bibinfo{author}{\bibfnamefont{M.~M.} \bibnamefont{Sala}},
  \bibinfo{author}{\bibfnamefont{V.}~\bibnamefont{Hinkov}},
  \bibinfo{author}{\bibfnamefont{M.~W.} \bibnamefont{Haverkort}},
  \bibinfo{author}{\bibfnamefont{M.}~\bibnamefont{Minola}},
  \bibinfo{author}{\bibfnamefont{M.}~\bibnamefont{Bakr}},
  \bibinfo{author}{\bibfnamefont{K.~J.} \bibnamefont{Zhou}},
  \bibinfo{author}{\bibfnamefont{S.}~\bibnamefont{Blanco-Canosa}},
  \bibnamefont{et~al.}, \bibinfo{journal}{Nat. Phys.}
  \textbf{\bibinfo{volume}{7}}, \bibinfo{pages}{725} (\bibinfo{year}{2011}).

\bibitem[{\citenamefont{Dean et~al.}(2013{\natexlab{b}})\citenamefont{Dean,
  Dellea, Springell, Yakhou-Harris, Kummer, Brookes, Liu, Sun, Strle, Schmitt
  et~al.}}]{dean2013persistence}
\bibinfo{author}{\bibfnamefont{M.~P.~M.} \bibnamefont{Dean}},
  \bibinfo{author}{\bibfnamefont{G.}~\bibnamefont{Dellea}},
  \bibinfo{author}{\bibfnamefont{R.~S.} \bibnamefont{Springell}},
  \bibinfo{author}{\bibfnamefont{F.}~\bibnamefont{Yakhou-Harris}},
  \bibinfo{author}{\bibfnamefont{K.}~\bibnamefont{Kummer}},
  \bibinfo{author}{\bibfnamefont{N.~B.} \bibnamefont{Brookes}},
  \bibinfo{author}{\bibfnamefont{X.}~\bibnamefont{Liu}},
  \bibinfo{author}{\bibfnamefont{Y.-J.} \bibnamefont{Sun}},
  \bibinfo{author}{\bibfnamefont{J.}~\bibnamefont{Strle}},
  \bibinfo{author}{\bibfnamefont{T.}~\bibnamefont{Schmitt}},
  \bibnamefont{et~al.}, \bibinfo{journal}{Nat. Mater.}
  (\bibinfo{year}{2013}{\natexlab{b}}).

\bibitem[{\citenamefont{Ellis et~al.}(2015)\citenamefont{Ellis, Huang,
  Olalde-Velasco, Dantz, Pelliciari, Drachuck, Ofer, Bazalitsky, Berger,
  Schmitt et~al.}}]{Ellis2015}
\bibinfo{author}{\bibfnamefont{D.~S.} \bibnamefont{Ellis}},
  \bibinfo{author}{\bibfnamefont{Y.-B.} \bibnamefont{Huang}},
  \bibinfo{author}{\bibfnamefont{P.}~\bibnamefont{Olalde-Velasco}},
  \bibinfo{author}{\bibfnamefont{M.}~\bibnamefont{Dantz}},
  \bibinfo{author}{\bibfnamefont{J.}~\bibnamefont{Pelliciari}},
  \bibinfo{author}{\bibfnamefont{G.}~\bibnamefont{Drachuck}},
  \bibinfo{author}{\bibfnamefont{R.}~\bibnamefont{Ofer}},
  \bibinfo{author}{\bibfnamefont{G.}~\bibnamefont{Bazalitsky}},
  \bibinfo{author}{\bibfnamefont{J.}~\bibnamefont{Berger}},
  \bibinfo{author}{\bibfnamefont{T.}~\bibnamefont{Schmitt}},
  \bibnamefont{et~al.}, \bibinfo{journal}{Phys. Rev. B}
  \textbf{\bibinfo{volume}{92}}, \bibinfo{pages}{104507}
  (\bibinfo{year}{2015}).

\end{thebibliography}
\end{document}